# The Spatial Dimension of Online Echo Chambers

Marco T. Bastos[1], Dan Mercea[1], and Andrea Baronchelli[2]

1. Department of Sociology - City, University of London
2. Department of Mathematics - City, University of London

**Abstract**

This study explores the geographic dependencies of echo-chamber communication on Twitter during the Brexit referendum campaign. We review the literature on filter bubbles, echo chambers, and polarization to test five hypotheses positing that echo-chamber communication is associated with homophily in the physical world, chiefly the geographic proximity between users advocating sides of the campaign. The results support the hypothesis that echo chambers in the Leave campaign are associated with geographic propinquity, whereas in the Remain campaign the reverse relationship was found. This study presents evidence that geographically proximate social enclaves interact with polarized political discussion where echo-chamber communication is observed. The article concludes with a discussion of these findings and the contribution to research on filter bubbles and echo chambers.

**Introduction**

Literature on online social networks defines echo chambers as a process of self-selection in which communication is circumscribed to ideologically-aligned cliques (Del Vicario et al., 2016; Del Vicario, Zollo, Caldarelli, Scala, & Quattrociocchi, 2017; Zollo et al., 2017). The political communication literature has explored the potential of echo chambers to foreclose deliberation by reinforcing the political views and preferences of individuals (Colleoni, Rozza, & Arvidsson, 2014; Wojcieszak & Mutz, 2009), a development largely seen as a problematic for democracy in as far as it engenders political polarization (Sunstein, 2007). Echo chambers are deeply embedded in processes of political polarization, selective exposure, and "filter bubbles," but research exploring the structural equivalence (Wasserman & Faust, 1994) of actors involved in echo-chamber communication remains largely uncharted (Barberá, Jost, Nagler, Tucker, & Bonneau, 2015).

While exact equivalence is rarely observed in real-world social networks, structural, isomorphic, and regular equivalence are measures of network similarity and equivalence that can help identifying classes or clusters of users (Freeman, 2011). Transposed to the network of tweets about the U.K. E.U. membership referendum, we seek to explore whether users engaging in echo-chamber communication are clustered in subgraphs that are homogeneous with regard to sociodemographic variables such as household and employment. More specifically, we consider whether geography and class are variables that interact with echo-chamber communication, a proposition in line with the principle of homophily asserting that individuals are more likely to associate and form ties with similar others. Isomorphic dependencies affecting homophily include a range of factors such as age, gender, class, and organizational role (Scott & Carrington, 2011), but for the purposes of this study we sought to extensively probe the relationship between geography and echo chambers.

Homophily thus offers a framework for understanding isomorphic dependencies that might interact with echo chamber communication. Firstly, echo chambers censor, disallow, or underrepresent competing views by enforcing social homogeneity much in line with the bandwagon effect predicted by the homophily model (Mark, 2003). Secondly, cultural similarities and differences among people can be formalized as a function of geographic propinquity (McPherson, Smith-Lovin, & Cook, 2001). Thirdly, online social networks are more prone to homophily compared with offline networks, which are tied to physical locations where serendipitous exposure to social diversity is more likely to happen (Hampton & Gupta, 2008; Hampton, Livio, & Goulet, 2010). In view of that, we seek to advance cognate scholarship with an investigation into whether echo-chamber communication is associated with geographical proximity between users that tweeted the U.K. E.U. membership referendum or the "Brexit campaign."

The Brexit referendum campaign was held at a time of sharp polarization among the electorate on the cultural and economic costs or, alternatively, the benefits of E.U. membership (Kriesi & Frey, 2008). While older and culturally conservative voters protested the infringement on national sovereignty by European institutions, in addition to being concerned about the impact of E.U. workforce mobility on British society, a more liberal-minded electorate welcomed the collective decision-making at the heart of the European Union and accepted the inward E.U. migration into the U.K. as a source of greater diversity (Curtice, 2016). In the run-up to the referendum, the winning campaign to leave the Union foregrounded a culturally conservative message centered on the proposition to 'take back control' from the E.U. by reasserting the sovereignty of the British Parliament and courts of justice and curtailing the free movement of E.U. labor (Leave, 2016; Leave.EU, 2016). The vote to leave the E.U. further exposed a geographical splintering of the country. The metropolitan area of London, Scotland and Northern Ireland voted to remain while the rest of England as well as Wales voted to leave (Rennie Short, 2016).

In what follows, we review the literature on political polarization and online echo chambers, chiefly diagnosed as a phenomenon further fragmenting the electorate in liberal democracies. We propose several hypotheses designed to advance the analysis of echo chambers by emphasizing their relationship to physical distance. Our results move forward this scholarship by foregrounding the geographical dependencies of Twitter echo chambers.

**Previous Work**

Polarization is a malaise of democratic politics proliferating through negative campaigning predicated on the partisan articulation of dismissive out-group stereotypes (Iyengar, Sood, & Lelkes, 2012). The factors driving political polarization have been thoroughly documented in the literature on information diffusion and they partially overlap with the factors driving homophily (Moody & Mucha, 2013). These factors include the tendency to expose oneself to consonant content that reinforces predispositions, particularly one's pre-existing political views (Iyengar & Hahn, 2009; Stroud, 2008, p. 360). Polarization also pertains to a pattern of association into opinion or attitude-based groups through a process of separation by disagreement (Baldassarri & Gelman, 2008; DiMaggio, Evans, & Bryson, 1996). As the distance between people's opinions widens, they become less likely to influence each other and more susceptible to clustering together in isolated groupings (Bessi et al., 2015). As polarization rises, cooperation to reach cross-ideological and agreeable outcomes becomes increasingly less likely (Andris et al., 2015).

The metaphor of echo chambers was traditionally used to caution against the rhetoric exalting the greater diversity and pluralism of the internet (Sunstein, 2007). The distributed infrastructure of the internet—spanning divisions of time and space (Castells, 2009)—was envisioned as enabling horizontal, networked congregation around mutual interests or practices that bypassed traditional intermediaries in modern democratic societies. The circumvention of intermediaries was, in the eyes of some observers, a threat to society as it removed a common framework for social experience (Sunstein, 2007). This latter set of concerns has however misrepresented the role of online interactions in the information flow from news-makers to audiences, particularly the homogenizing effect of inter-personal relationships on the opinions and actions of individuals. Likewise, it fails to take account of the evidence that the media have embraced and amplified negative campaigning by political candidates (Castells, 2009; Iyengar, et al., 2012).

Notwithstanding, the individual seemingly became at once embedded in multiple, personal and diverse social networks thanks to the usage of internet technologies (Rainie & Wellman, 2012, p. 7). This networked individual supposedly substituted membership in dense social groups that are strongly tied to one's physical location—e.g., of one's home, neighborhood, or workplace—with immersion into geographically dispersed online networks of shared interest, practice, or outlook (Rainie & Wellman, 2012; Sunstein, 2007; Wojcieszak & Mutz, 2009). This levelling of geographical distance did not prevent individuals from grouping with similar others online (McPherson, et al., 2001), but the relationship between spatial distance and users' interaction in social media has been found to be significant, with friendship ties in densely connected groups arising at shorter spatial distances compared with social ties between members of different groups (Laniado, Volkovich, Scellato, Mascolo, & Kaltenbrunner, 2017). More importantly, research found social ties on Twitter to be constrained by geographical distance with an over-representation of ties confined to distances shorter than 100 kilometers (Takhteyev, Gruzd, & Wellman, 2012).

If anything, networked individualism complicated social interaction in two notable ways. First, the use specifically of social media was shown to be coextensive with larger and more diverse personal networks likely to include a discussion confidant from a different political party (Hampton, Sessions, & Her, 2011). Hampton, et al. (2010) suggested that a pervasive awareness of one's own network resulted from social networking sites making one's own ego networks publicly visible. Subsequently, the commercial services underpinning social media platforms deployed algorithms designed to quantify and monetize social interaction (Dijck, 2013), narrowly confining it to a filter bubble algorithmically populated with information closely matching observed and expressed user preferences (Pariser, 2012, p. 11).

More than a decade of research into echo chambers, however, suggests that the use of internet technologies was not an immediate cause of selective exposure or ideological polarization, the latter previously appearing as more pronounced in face-to-face interactions (Gentzkow & Shapiro, 2011). Notably, Horrigan, Garrett, and Resnick (2004, p. ii) found that "internet users are not insulating themselves in information echo chambers. Instead, they are exposed to more political arguments than non-users". Users of algorithmically personalized filtering systems seemed similarly exposed to higher numbers of sources, channels and news categories (Beam & Kosicki, 2014). Indeed, exposure to diverse and even competing opinions on polarizing topics was found to occur on social media platforms across various national contexts (Bakshy, Messing, & Adamic, 2015; Fletcher & Nielsen, 2017; Kim, 2011).

While researchers have not found filter bubbles to be associated with news readership, evidence of echo-chamber communication and filter bubbles was found in other settings, with representative surveys identifying social media users who actively seek the company of people with whom they tend to agree (Vaccari et al., 2016). Notwithstanding the rapid expansion of online social networks, and the attendant expectation of higher exposure to a variety of news and politically diverse information (Messing & Westwood, 2014, p. 1056), online social networks also increased the appetite for selective exposure in highly polarized social environments (Wojcieszak, 2010), with the sharing of controversial news items being particularly unlikely to take place in these contexts (Bright, 2016). Even with scanty evidence linking filter bubbles and echo chambers to general social media communication, researchers have found evidence of echo-chamber communication in polarized contexts and balkanized media landscapes (Barberá, et al., 2015; Iyengar & Hahn, 2009; Stroud, 2008; Wojcieszak & Mutz, 2009).

Barberá, et al. (2015) emphasized that political information was more likely to be retweeted if received from ideologically similar sources, a finding that lends support to the selective exposure hypothesis. Song and Boomgaarden (2017) explored selective exposure resulting from the choice of information sources being aligned with political preferences. Himelboim, McCreery, and Smith (2013) reported that cross-ideological information on Twitter was unlikely to be circulated in social clusters with a strong group identity. These results speak to early research on selective exposure on social media that found politically conservative users to be more likely than liberals to confine their communication of political issues to ideological echo chambers on blogs (Adamic & Glance, 2005), Twitter (Adamic & Glance, 2005; Barberá, et al., 2015; Hayat & Samuel-Azran, 2017) and Facebook (Mitchell & Weisel, 2014).

Inversely, liberals have tended to engage in cross-ideological retweeting more systematically than conservatives (Barberá, et al., 2015) while clustering around partisan information sources is more prevalent among conservative than liberal voters (Benkler, Faris, Roberts, & Zuckerman, 2017). The use of "moral-emotional language" on Twitter was found to be associated with polarization and to lead to echo chambers (Brady, Wills, Jost, Tucker, & Van Bavel, 2017) while @-mentioning was found to be associated with higher degrees of cross-ideological interactions (Conover et al., 2011), likely as a strategy to tinge rival political content with partisan messages. In the last instance, however, bridging echo chambers is a cost-intensive activity that requires engaging in a combative discussion and disputing competing claims (Bessi, et al., 2015; Bessi et al., 2016).

**Hypotheses**

Taking these considerations into account, we expect echo-chamber communication to be prevalent in the referendum campaign (H1)—i.e., that users identified with one side of the campaign will be more likely to engage with like-minded users. We refer to this pattern of echo chamber in which both sender and receiver are tweeting the same campaign as *in-bubble* interactions and we test this hypothesis for the Leave (H1a) and the Remain (H1b) campaigns. In view of the homophily dependencies reviewed hitherto, particularly the association between selective exposure and social enclaves (social groupings with specific interests within larger groups), we also expect echo-chamber communication to be at least partially mirrored by geographic proximity. Therefore, we hypothesize that echo chambers are to be found predominantly in geographically proximate *in-bubbles* both in the Leave (H2a) and the Remain (H2b) campaigns. We refer to this geographically-bounded self-selection as *neighboring in-bubble* communication in opposition to *cross-bubble*, when users interact with the other side of the campaign, and *out-bubble*, when users interact with a neutral user.

Thirdly, we take stock of the geographic factor possibly driving echo chambers and hypothesize that *in-bubbles* will cover shorter geographic distances compared with *out-* and *cross-bubbles*, both in the Leave (H3a) and Remain (H3b) campaigns. Hypotheses H2 and H3 are thus intrinsically connected: while the former explores whether geographically-proximate echo chambers are prevalent, the latter (H3) probes whether echo-chamber communication is more likely to cover short geographic distances. Testing these hypotheses allows us to subsequently probe whether Leave interactions are more likely to be neighboring in-bubbles compared with Remain interactions (H4). Lastly, and most importantly for the aims of this study, we postulated that echo-chamber communication (*in-bubble*) is associated with the geographic distance covered by messages in the Leave (H5a) and Remain (H5b) campaigns. In summary, we proceed from a relatively simple hypothesis of whether echo-chamber communication occurred in the Brexit debate on Twitter towards the last hypothesis testing the geographic dependencies of echo chambers. As such, the five main hypotheses to be tested in this study are the following:

*H1a.*   Users tweeting the Leave campaign interact predominantly with other users tweeting the Leave campaign (echo chambers within the Leave campaign)

*H1b.*   Users tweeting the Remain campaign interact predominantly with other users tweeting the Remain campaign (echo chambers within the Remain campaign)

*H2a.*   Leave interactions are predominantly within geographically proximate echo chambers (neighboring in-bubble)

*H2b.*   Remain interactions are predominantly within geographically proximate echo chambers (neighboring in-bubble)

*H3a.*   Leave echo chambers cover shorter geographic distances compared with non-bubbles

*H3b.*   Remain echo chambers cover shorter geographic distances compared with non-bubbles

*H4.*   Leave interactions are more likely to be neighboring in-bubbles compared with Remain interactions

*H5a.*   Echo-chamber communication in the Leave campaign is associated with geographic distance

*H5b.*   Echo-chamber communication in the Remain campaign is associated with geographic distance

**Data and Methods**

For the purposes of this project we relied on Twitter Streaming and REST APIs to collect a total of 5,099,180 tweets using a set of keywords and hashtags, including relatively neutral tags such as referendum, inorout, and euref, but more importantly, messages that used hashtags clearly aligned with the Leave campaign: voteleave, leaveeu, takecontrol, no2eu, betteroffout, voteout, britainout, beleave, iwantout, and loveeuropeleaveeu; and hashtags clearly aligned with the Remain campaign: strongerin, leadnotleave, votein, voteremain, moreincommon, yes2eu, yestoeu, betteroffin, ukineu, and lovenotleave. The campaign-aligned, hashtag-based datasets are leveraged to identify messages advocating each side of the referendum: the Vote Leave or Vote Remain campaign. We subsequently removed messages tweeted before 15 April 2016, the starting date of the official campaign period, and after 24 June 2016, the end of the referendum campaign.

Next we identify the location of users in our dataset by triangulating information from geocoded tweets (subsequently reverse-geocoded), locations identified in their user profile (then geocoded), and information that appeared in their tweets. The triangulation prioritizes the signal with higher precision, hence geocoded information is preferred if present. When not available, we look at the location field in users' profiles and geocode that location. If neither source of information is available, we check for information in their tweets, but only in cases where the *place_id* field of the API response returns relevant information. As a result, a considerable portion of user locations in our dataset could be identified only to city or postcode level. Upon identifying the location of users, we remove users located outside the United Kingdom or whose location we could not identify up to postcode level. This reduces our dataset to 565,028 messages or 11% of all collected messages; a sample of messages that is sufficiently large to allow for exploring the geographic dispersion of Vote Leave and Vote Remain Campaigns.

For each tweet, we count the number of hashtags advocating the Leave and Remain campaigns. We tag the message as Remainer or Leaver on the basis of the highest number of hashtags used. Messages without hashtags advocating either side of the campaign are tagged as Neutral. This information, once aggregated, is also used to identify the affiliation of users that tweeted or retweeted politically polarized hashtags. Highly polarized messages—i.e., tweets including several one-sided hashtags, are however uncommon. For users championing the Vote Leave campaign, only 16% of their messages included more than one one-sided hashtag. These messages are yet more uncommon in the vote Remain campaign, where only 2% of messages included more than one hashtag clearly associated with that side of the campaign.

We subsequently identify the campaign affiliation (Leave or Remain) of users @-mentioned or retweeted in the original tweet. To achieve this, we loop through the dataset to find messages tweeted by these recipients that championed either side of the campaign. We calculate the mode or "mean affiliation" per user based on the frequency of one-sided hashtags used throughout the period. The mean affiliation per user can only be calculated for users that actively participated in the referendum campaign on Twitter. In other words, for users at the receiving end (@-mentioned or retweeted) to be identified as Leaver or Remainer, the user in question must have tweeted or retweeted a separate tweet with hashtags clearly aligned with one side of the campaign, whereas users that tweeted equal number of Remain and Leave hashtags are tagged as neutral. The rationale for restricting the parameters of ideological identification between users was to avoid mainstream media and high-profile accounts, which are regularly @-mentioned or whose tweets are retweeted with the addition of one-sided hashtags, to be classified in either side of the campaign battle. The mean affiliation has the added benefit of filtering out retweets or @-mentions intended as provocation or ironic remark; these messages are offset by the broader ideological orientation tweeted by the account, and users that have only sourced information or received @-mentions are classified as neutral for not having themselves tweeted any partisan hashtag.

In short, we opted for a more conservative approach to identifying campaign affiliation at the receiving end of a tweet so that users are only associated with one side of the campaign if the user herself tweeted a partisan message at some point during the campaign. We believe this approach grounded on the mean affiliation per user reflects strong campaign membership with low probability of false-positives, but these conservative parameters to identifying campaign affiliation further reduced our dataset to 33,889 tweets. The multiple sampling of the data (timespan, geographic location, campaign affiliation of sender and receiver) rendered a dataset that is both easy to process and highly curated by geographically enriched data. Given the rationale of this project, we believe this dataset offers a defensible if limited representation of the debate and our conclusions are conditional on these constraints.

Echo chambers are defined as a function of the identified campaign affiliation. We tag each tweet as in-bubble if sender and receiver (@-mentioned or retweeted) have tweeted the same campaign. We tag it as cross-bubble if the sender has tweeted one campaign and the receiver (@-mentioned or retweeted) has tweeted the opposite campaign. We tag the tweet as out-bubble if either sender or receiver was classified as neutral, which simply means any of them have not tweeted any message with clearly supportive campaign hashtags. We further identified 237 users with suspicious bot activity, but whose echo chamber activity comprised of only 63 messages. To control for potential issues associated with bot activity, we replicated the analysis without this group of users, but no difference was found in the results.

With the location of users defined using the abovementioned triangulation approach, we leverage the longitude and latitude values to calculate the Euclidean distance (in kilometers) covered by the sender and receiver of @-messages and retweets. We use the canonical mean equatorial radius (6378.145 km or 2.092567257E7 ft.) for earth radius. As such, our calculation is not mathematically precise due to the inaccurate estimate of the earth's radius (R). Despite this perennial limitation, we believe the calculation is adequate as mathematical precision is of lesser importance when analyzing data whose geographic accuracy is limited to postcode level. We repeat the process for each tweet, thus identifying the account being @-mentioned or retweeted and calculating the distance (in kilometers) between sender and receiver.

Finally, these differences are analyzed with a series of statistical tests, including linear regression, Chi-square, and Kolmogorov–Smirnov. For the Chi-squared tests, we reject the null hypothesis of the independence assumption if the *p*-value of $x^2 = \sum_{i,j} \frac{(f_{i,j} - e_{i,j})^2}{e_{i,j}}$ is less than the given significance level $\alpha$. To test Hypotheses H5, we examine if the variables sender's affiliation and receivers' affiliation are independent and if the probability distribution of one variable is affected by the other.

**Results**

We evaluated H1 and H2 by testing whether the probability distribution of the sender's and receiver's affiliation are independent (i.e., one variable is not affected by the presence of another). The variables are significantly correlated (r=.66, *p*<.0001) with a Chi-squared value of 9646.4 (*p*< 2.2e-16). As a result, we conclude that the two variables are in fact dependent. We further explored H1a and rejected the null hypothesis of the independence assumption at the 95% confidence level, as users tweeting the Leave campaign are significantly more likely to interact with users also tweeting highly partisan Leave hashtags. In fact, only 9% of messages tweeted by users affiliated with the Leave campaign were directed to users associated with the Remain campaign (cross-bubble communication). This contrasted with 22% of interactions directed to neutral users (out-bubble communication) and a towering 69% of Leave @-messages and retweets being sourced from or directed to another user affiliated with the Leave campaign (in-bubble communication), with little to no difference between @-mentions and retweets.

The intensity of echo-chamber communication is remarkably similar on the Remain side of the campaign (H1b), where only 10% of users directed @-mentions or retweeted content from users identified with the Leave campaign (cross-bubble communication), with 22% of interactions including neutral users (out-bubble communication), and a total of 68% of interactions initiated by Remainers being echo chambers (in-bubble communication). The likelihood of users campaigning for one side of the referendum engaging with users of the same leaning—instead of neutral or adversarial users—was captured by fitting a linear regression model on the sender's affiliation as the explanatory variable of echo-chamber communication: partisan affiliation explains nearly half of the variance in the data ($R^2_{adj}$=.44, $p$<2.2e-16). Figure 1 unpacks these findings and shows the prevailing patterns of echo-chamber communication compared with out-bubble and cross-bubble communication (complementarity), both in the Leave and the Remain campaign, across a range of distance radiuses.

FIGURE 1 ABOUT HERE

We approached H2 by examining whether Leave and Remain interactions are predominantly within neighboring in-bubbles or geographically proximate echo chambers—i.e., within a 50 kilometers radius expanded in 50 kilometers increments up to 900 kilometers, which is the maximum straight-line distance between two geographical points in the United Kingdom (from Land's End to John o' Groats). We found that most interactions are within a 200km radius, but the geographic trend is different between the Leave and Remain campaigns. As shown in Figure 1a, the Cumulative Distribution Function (CDF) of in-bubble Leave messages covers shorter distances compared to non-bubbles (i.e., out- and cross-bubble), with half of in-bubble messages covering less than 200 kilometers. The trend is reversed on the Remain side of the campaign, in which in-bubble interactions cover longer distances compared to non-bubble messages.

Figure 1b also shows that Leave-campaign messages are chiefly exchanged within ideologically and geographically proximate echo chambers, a component of echo-chamber communication behavior that we further uncover in the following analyses. There is also relatively little cross-ideological retweeting and @-mentioning, much in line with previous findings reported in the literature (Himelboim, et al., 2013). The trend is however inversed on the Remain side of the campaign: as distance between sender and receiver increases, in-bubble communication becomes more common and covers increasingly larger geographic areas compared to out- and cross-bubble interactions. This reversed trend depicted in the CDF plot is also captured by the mean distance covered by Leave messages, at 199km for in-bubble and 234km for non-bubble ($\tilde{x}$=168 and $\tilde{x}$=208, respectively). For Remain messages, contrariwise, the mean distance is 238km for in-bubble and 204km for non-bubble ($\tilde{x}$=209 and $\tilde{x}$=184, respectively).

This is consistent with Hypothesis H2a and H2b, which state that Leave and Remain interactions, respectively, are predominantly within neighboring in-bubbles. Figure 2 shows this relationship by contrasting cross-, out-, and in-bubble communication across the United Kingdom (we found no difference in the communication patterns of @-mentions and retweets). The results lend support to H3a but reject hypothesis H3b, as only Leave echo chambers are likely to cover short geographic distances compared with non-bubbles. In fact, there are nearly three times as many in-bubble interactions in the Leave campaign for every cross-bubble and out-bubble communication combined, with the number of users involved in out-bubble communication being about half and one-fifth, respectively, of those involved in echo-chamber communication.

Despite echo chambers on the Remain side being independent from geographic distance, there is a much higher ratio of interactions falling within in-bubble patterns. Similarly to the Leave campaign, there are three times as many in-bubble interactions for every out-bubble, and six times as many for every cross-bubble interaction. The number of Remain supporters involved in out-bubble communication is about two-thirds of those involved in in-bubble communication and only a third if we compare cross-bubble with in-bubble. However, while echo chambers in the Leave campaign appear constrained by short geographic distances (H3a), this is not the case on Remain side (H3b). In fact, Remain echo chambers are likely to span greater geographic distances while their cross-bubble communication is concentrated around neighboring communities.

FIGURE 2 ABOUT HERE

We approach H4, which hypothesizes that Leave interactions cover shorter geographic distances compared with Remain interactions, by calculating the average distance @-mentions and retweets travelled from source to destination for each side of the partisan divide. The results lend support to H4 as one-quarter of Leave interactions took place within 100 kilometers whereas fewer than one-fifth followed such pattern in the Remain campaign. The average distance covered by Leave partisan messages is also shorter at 178km compared with 199km for the Remain campaign. In-bubble communication in the Leave campaign covers a remarkably short geographic distance at only 22 kilometers compared to 40km for the Remain campaign, a pattern also observed in out-bubble and cross-bubble communication, where Leave messages cover 86 and 197 kilometers compared with 103 and 243 for Remain messages, respectively. Figure 2 unpacks these differences and shows the geographic clustering of Leave messages, particularly in-bubble interactions, centered in the Brexit heartland of the English Midlands, the North, and the East.

      We approach H5a, which hypothesizes that echo-chamber communication is associated with geographic proximity in the Leave and Remain campaigns (H5a and H5b, respectively), by comparing the density distribution curves of in-, out-, and cross-bubble communication subgraphs alongside the density curve of randomly-generated comparable subgraphs. To this end, we randomly swap the location of users in each subgraph (in-, out-, and cross-bubble), recalculate the distance travelled by @-mention and retweet messages, and compare the observed distribution of distances against the random distribution of distances travelled by each message. The rationale for this analysis is to identify distributions that deviate from the random reallocation of users across geographic locations while preserving individual social networks identified by their communications on Twitter as well as the geographic distribution of users in the country. We do not assign random locations to users; we simply swap the location of users in each subgraph to test if the distribution is similar to the random network which preserves the overall geographical distribution of users. This approach establishes an association between echo-chamber communication and the geography of message diffusion whenever the observed networks—*ceteris paribus*—differ significantly from the random network. In other words, for each iteration of the test we retain the set of locations in each subgraph, but randomly reorder the

locations to test whether geographic dependencies found in echo-chamber communication are replicated in the randomized geographic network.

We ran 100 iterations of each test and the results are shown in Figure 3: the high volume of interactions within geographically proximate echo chambers—i.e., within the 50 kilometers radius—is a considerable departure from the distribution in the randomized network. This deviation is particularly prominent in echo-chamber communication (i.e., in-bubble interactions). This pattern disappears when the location of users is randomly reshuffled, an indication that the distribution is not determined by chance. We can thus conclude that the geographic distribution of echo-chamber communication is unlikely, i.e., much less likely to happen than in the randomized null model. This unlikely distribution is yet more salient in the subgraph of Leave, in-bubble interactions and disappears in out- and cross-bubble interactions for the Leave campaign and again in the entire network of Remain interactions. In other words, the association between geographic proximity and echo-chamber communication is restricted to the Leave campaign and lends support to hypothesis H5a, while in the Remain campaign we observe no such dynamics and hence reject hypothesis H5b.

FIGURE 3 ABOUT HERE

To assess the significance of these results, we performed a Kolmogorov–Smirnov test on the probability distribution of distances covered by messages compared with the probability distribution of distances covered by messages with users' locations randomly reshuffled. In other words, we swap the location of users in the graph and calculate the distances covered by their interactions again, thus providing a reference probability distribution to test the similitude of the two samples with a continuous distribution. Figure 3 shows the test statistic, the maximum distance between the ECDF of the two samples, and the *p*-value for each of the tests. The results are significant for all modalities of self-selected bubble and each side of the political divide, except for cross-bubble communication, which is not significant in any of the subgraphs, likely a result of the small sample size of cross-bubble communication as the distributions are similar with no superimposed oscillatory disturbances.

The maximum distance (supremum) between the CDFs of the two samples is significantly higher for Leave in-bubble interactions, in which the peak amplitude deviates from the pattern observed for the rest of the network and during the random reshuffling of users' locations. The probability of seeing a test statistic as high or higher than the one observed if the two samples were drawn from the same distribution is vanishingly small. The results thus support hypothesis H5a: echo-chamber communication in the Leave campaign is likely to be associated with geographic proximity, as in-bubble interactions in the Leave campaign show significantly more short distance activity than expected by chance ($p<2.20E-16$). Hypothesis H5b—that echo-chamber communication is associated with geographic proximity in the Remain campaign—is however rejected. Although the distribution of in-bubble data for the Remain campaign deviates significantly ($p<1.56E-05$) from the reference probability distribution (stochastic rearrangement of users' locations), the effect is in effect inversed: echo-chamber communication in the Remain campaign is more likely to cover larger distances compared with out- and cross-bubble communication, which on average cover shorter geographic distances.

In view of the high deviation from the probability distribution of users' locations randomly reshuffled, we sought to further examine hypothesis H5 by probing variables that could have interfered with this distribution. We firstly speculated that highly-active, super-users in few cities could have drawn the geographic distribution of in-bubble communication in the Leave campaign. Secondly, we conjectured that isolated events such as the murder of the Labour Member of Parliament Jo Cox could have likewise skewed a distribution that would otherwise remain comparable to the remainder of the campaign data. However, we managed to rule out these effects by inspecting the probability distribution while controlling for these variables. To this end, we performed Kolmogorov–Smirnov tests on the observed subgraph of echo-chamber communication in the Leave campaign and the same subgraphs minus super-users (maximum 10 tweets). The results rejected the hypothesis that the two distributions are significantly different at the 95% confidence interval and supports the assumption that the geographic patterning found in echo chambers is independent of super-users.

FIGURE 4 ABOUT HERE

We addressed our second conjecture by slicing the 10-week period covered by the referendum campaign (14 April to 23 June) into four sub-periods comprising weeks 1-3, weeks 4-6, weeks 7-8, and weeks 9-10 and performing Kolmogorov–Smirnov tests on each temporal scenario. The distribution appears to change over time, but the geographic patterning associated with echo chambers in the Leave campaign remains relatively robust throughout the period. Figure 4 shows the observed and random distribution for echo chambers in the referendum network and subgraphs of the Leave and Remain campaigns over the 10-week period. Weeks 1-3 show a similar peak observed in the aggregate network, which decreases in the weeks 4-8, but surges again the last two weeks of the referendum, which concentrates most of the user activity in the period. It is interesting to note that in weeks 9-10, which is marked by the intense activity of Leave and Remain campaigners, the observed distribution is remarkably similar to the randomized signal at the network level, but the separate inspection of Leave and Remain subgraphs reveals striking interactions between online activity and geography.

In summary, the weekly variations continued to present peak amplitudes that deviate from the rest of the network and from the distribution observed with the random reshuffling of users' locations, with inverse geographic patterning of echo chambers for Leave and Remain campaigns, particularly in the weeks leading up to the vote. Therefore, we conclude that echo chambers in the Leave campaign are significantly associated with geographic propinquity and the results appear to be robust across classes of Twitter users and during different moments of the referendum campaign.

**Discussion and Conclusion**

This study presented the first evidence that geographically proximate social enclaves interact with polarized political discussions in which online echo-chamber communication is observed. The first hypothesis tested in this study is broadly consistent with previous results found in the literature (Del Vicario, et al., 2016; Zollo, et al., 2017). Del Vicario, et al. (2017) studied the Brexit debate on Facebook and reported significant evidence of echo chamber behavior. The finding that Remainers were more likely to engage in cross-bubble is also consistent with previous research that found liberals more likely to engage in cross-ideological retweeting than conservatives (Barberá & Rivero, 2015). While research on filter bubbles cautions against the inclination to identify filter bubbles with the medium of the internet (Fletcher & Nielsen, 2017), literature exploring political polarization on social media posits a positive prospect of echo-chamber communication being observed as the debate becomes more contentious (Barberá, et al., 2015). The results of hypothesis H1 are thus broadly consistent with previous research: the

Brexit debate has accentuated the political divides among the British public along antagonistic fault lines. As such, it is unsurprising to have found Leave and Remain campaigns engaging in widespread echo chamber behavior.

It is the results of hypotheses H2-H5 that shed new light on the dynamics of echo-chamber communication. By identifying a physical patterning in echo-chamber communication, at least in the Leave campaign, where in-bubble communication was remarkably restricted to physical communities proximate to each other, we have explored one isomorphic dependency underlying echo chambers—geographic propinquity—, but we nonetheless expect others to be at play, likely interacting with the geographic clustering in social and political enclaves that may have marked much of the Brexit debate. The findings also provide more granularity for future research exploring echo chambers: after extricating echo chambers from filter bubbles, we advance a geographical variable interacting with this pattern of communication that needs to be further investigated, if possible replicated, by future research. One compelling reason for replication is the notable claim that a tendency towards homophilic political clustering may be compounded by the geographic unboundedness of online communication (Wojcieszak & Mutz, 2009).

Hypothesis H2 offers the added benefit of working as a control mechanism. We previously established that echo-chamber communication was prevalent in the Brexit debate, and that in-bubble interactions were more likely to cover short geographic distances for the Leave campaign even after controlling for highly-active users and seasonal variations. Yet only by inspecting the entire network, along with out- and cross-bubble subgraphs, one could identify the geographic dependency as a development restricted to in-bubble, echo-chamber communication. We believe this is an important contribution to the debate on echo chambers, but further research is necessary to establish the magnitude of the effect and the underlying mechanisms through which physical proximity and political affinity translates into in-bubble, echo-chamber communication. We expect more intricate relationships between existing physical ties and online interactions to be at play (Laniado, et al., 2017; Takhteyev, et al., 2012), with the relationship between campaign affiliation and geographical propinquity capturing only secondary effects of this interaction. Ultimately, the results reported in this study are particularly puzzling when considering that the geographic embedding of online echo chambers was restricted to the Leave campaign. In sharp contrast, Remainers appear to have focused their cross-campaign efforts on neighboring areas of their community while their echo chambers aimed more distant, likely less tangible user accounts such as media outlets.

References

Adamic, L., & Glance, N. (2005). *The political blogosphere and the 2004 U.S. election: Divided they blog.* Paper presented at the 3rd International Workshop on Link Discovery (LinkKDD).

Andris, C., Lee, D., Hamilton, M. J., Martino, M., Gunning, C. E., & Selden, J. A. (2015). The Rise of Partisanship and Super-Cooperators in the U.S. House of Representatives. *PLoS ONE, 10*(4), e0123507. doi: 10.1371/journal.pone.0123507

Bakshy, E., Messing, S., & Adamic, L. (2015). Exposure to ideologically diverse news and opinion on Facebook. *Science, 348*(6239), 1130-1132. doi: 10.1126/science.aaa1160

Baldassarri, D., & Gelman, A. (2008). Partisans without Constraint: Political Polarization and Trends in American Public Opinion. *American Journal of Sociology, 114*(2), 408-446. doi: 10.1086/590649

Barberá, P., Jost, J. T., Nagler, J., Tucker, J. A., & Bonneau, R. (2015). Tweeting From Left to Right: Is online political communication more than an echo chamber? *Psychological Science, 26*(10), 1531-1542. doi: 10.1177/0956797615594620

Barberá, P., & Rivero, G. (2015). Understanding the Political Representativeness of Twitter Users. *Social Science Computer Review, 33*(6), 712-729. doi: doi:10.1177/0894439314558836

Beam, M. A., & Kosicki, G. M. (2014). Personalized News Portals: Filtering Systems and Increased News Exposure. *Journalism & Mass Communication Quarterly, 91*(1), 59-77. doi: 10.1177/1077699013514411

Benkler, Y., Faris, R., Roberts, H., & Zuckerman, E. (2017, 3 March 2017). Breitbart-led right-wing media ecosystem altered broader media agenda, *Columbia Journalism Review*. Retrieved from http://www.cjr.org/analysis/breitbart-media-trump-harvard-study.php

Bessi, A., Petroni, F., Vicario, M. D., Zollo, F., Anagnostopoulos, A., Scala, A., et al. (2015). *Viral Misinformation: The Role of Homophily and Polarization.* Paper presented at the Proceedings of the 24th International Conference on World Wide Web, Florence, Italy.

Bessi, A., Zollo, F., Del Vicario, M., Puliga, M., Scala, A., Caldarelli, G., et al. (2016). Users Polarization on Facebook and Youtube. *PLOS ONE, 11*(8), e0159641. doi: 10.1371/journal.pone.0159641

Brady, W. J., Wills, J. A., Jost, J. T., Tucker, J. A., & Van Bavel, J. J. (2017). Emotion shapes the diffusion of moralized content in social networks. *Proceedings of the National Academy of Sciences, 114*(28), 7313-7318. doi: 10.1073/pnas.1618923114

Bright, J. (2016). The social news gap: how news reading and news sharing diverge. *Journal of Communication, 66*(3), 343-365.

Castells, M. (2009). *The rise of the network society* (New ed.). Oxford: Wiley-Blackwell.

Colleoni, E., Rozza, A., & Arvidsson, A. (2014). Echo Chamber or Public Sphere? Predicting Political Orientation and Measuring Political Homophily in Twitter Using Big Data. *Journal of Communication, 64*(2), 317-332. doi: 10.1111/jcom.12084

Conover, M. D., Ratkiewicz, J., Francisco, M., Goncalves, B., Menczer, F., & Flammini, A. (2011). *Political Polarization on Twitter*.

Curtice, J. (2016). A Question of Culture or Economics? Public Attitudes to the European Union in Britain. *The Political Quarterly, 87*(2), 209-218. doi: 10.1111/1467-923X.12250

Del Vicario, M., Bessi, A., Zollo, F., Petroni, F., Scala, A., Caldarelli, G., et al. (2016). The spreading of misinformation online. *Proceedings of the National Academy of Sciences, 113*(3), 554-559. doi: 10.1073/pnas.1517441113

Del Vicario, M., Zollo, F., Caldarelli, G., Scala, A., & Quattrociocchi, W. (2017). Mapping social dynamics on Facebook: The Brexit debate. *Social Networks, 50*, 6-16. doi: http://dx.doi.org/10.1016/j.socnet.2017.02.002

Dijck, v., J. A. (2013). *The Culture of Connectivity: A Critical History of Social Media*. Oxford: Oxford University Press.

DiMaggio, P., Evans, J., & Bryson, B. (1996). Have American's Social Attitudes Become More Polarized? *American Journal of Sociology, 102*(3), 690-755.

Fletcher, R., & Nielsen, R. K. (2017). Are News Audiences Increasingly Fragmented? A Cross-National Comparative Analysis of Cross-Platform News Audience Fragmentation and Duplication. *Journal of Communication*. doi: 10.1111/jcom.12315

Freeman, L. C. (2011). The Development of Social Network Analysis – with an Emphasis on Recent Events. In J. Scott & P. J. Carrington (Eds.), *The SAGE handbook of social network analysis*. London: SAGE publications.

Gentzkow, M., & Shapiro, J. M. (2011). Ideological Segregation Online and Offline. *The Quarterly Journal of Economics, 126*(4), 1799-1839.

Hampton, K. N., & Gupta, N. (2008). Community and social interaction in the wireless city: wi-fi use in public and semi-public spaces. *New Media & Society, 10*(6), 831-850. doi: 10.1177/1461444808096247

Hampton, K. N., Livio, O., & Goulet, L. (2010). The social life of wireless urban spaces : internet use, social networks, and the public realm. *Journal of Communication, 60*(4), 701-722.

Hampton, K. N., Sessions, L. F., & Her, E. J. (2011). Core networks, social isolation and new media. *Information, Communication & Society, 14*(1), 130-155. doi: 10.1080/1369118X.2010.513417

Hayat, T., & Samuel-Azran, T. (2017). "You too, Second Screeners?" Second Screeners' Echo Chambers During the 2016 U.S. Elections Primaries. *Journal of Broadcasting & Electronic Media, 61*(2), 291-308. doi: 10.1080/08838151.2017.1309417

Himelboim, I., McCreery, S., & Smith, M. (2013). Birds of a Feather Tweet Together: Integrating Network and Content Analyses to Examine Cross-Ideology Exposure on Twitter. *Journal of Computer-Mediated Communication, 18*(2), 40-60. doi: 10.1111/jcc4.12001

Horrigan, J., Garrett, K., R., & Resnick, P. (2004). The internet and democratic debate *Pew Internet & American Life Project*. Washington, D.C.

Iyengar, S., & Hahn, K. S. (2009). Red Media, Blue Media: Evidence of Ideological Selectivity in Media Use. *Journal of Communication, 59*(1), 19-39. doi: 10.1111/j.1460-2466.2008.01402.x

Iyengar, S., Sood, G., & Lelkes, Y. (2012). Affect, Not IdeologyA Social Identity Perspective on Polarization. *Public Opinion Quarterly, 76*(3), 405-431. doi: 10.1093/poq/nfs038

Kim, Y. (2011). The contribution of social network sites to exposure to political difference: The relationships among SNSs, online political messaging, and exposure to cross-cutting perspectives. *Computers in Human Behavior, 27*(2), 971-977. doi: http://dx.doi.org/10.1016/j.chb.2010.12.001

Kriesi, H., & Frey, T. (2008). The United Kingdom: moving parties in a stable configuration. In E. Grande, H. Kriesi, M. Dolezal, R. Lachat, S. Bornschier & T. Frey (Eds.), *West European Politics in the Age of Globalization* (pp. 183-207). Cambridge: Cambridge University Press.

Laniado, D., Volkovich, Y., Scellato, S., Mascolo, C., & Kaltenbrunner, A. (2017). The Impact of Geographic Distance on Online Social Interactions. *Information Systems Frontiers*, 1-16.

Leave, V. (2016). Taking Back Control From Brussels, from http://www.voteleavetakecontrol.org/briefing_control.html

Leave.EU. (2016). Our Vision, from http://leave.eu/about/

Mark, N. P. (2003). Culture and competition: Homophily and distancing explanations for cultural niches. *American sociological review*, 319-345.

McPherson, M., Smith-Lovin, L., & Cook, J. M. (2001). Birds of a Feather: Homophily in Social Networks. *Annual Review of Sociology, 27*(1), 415-444. doi: 10.1146/annurev.soc.27.1.415

Messing, S., & Westwood, S. J. (2014). Selective Exposure in the Age of Social Media. *Communication Research, 41*(8), 1042-1063. doi: doi:10.1177/0093650212466406

Mitchell, A., & Weisel, R. (2014). Political polarization and media habits: Pew Research Center.

Moody, J., & Mucha, P. J. (2013). Portrait of Political Party Polarization. *Network Science, 1*(01), 119-121. doi: doi:10.1017/nws.2012.3

Pariser, E. (2012). *The filter bubble: what the internet is hiding from you*. London: Penguin.

Rainie, H., & Wellman, B. (2012). *Networked: the new social operating system*. Cambridge, MA: MIT Press.

Rennie Short, J. (2016). The Geography of Brexit: What the Vote Reveals about Disunited Kingdom, *The Conversation*. Retrieved from http://theconversation.com/the-geography-of-brexit-what-the-vote-reveals-about-the-disunited-kingdom-61633

Scott, J., & Carrington, P. J. (2011). *The SAGE handbook of social network analysis*. London: SAGE publications.

Song, H., & Boomgaarden, H. G. (2017). Dynamic Spirals Put to Test: An Agent-Based Model of Reinforcing Spirals Between Selective Exposure, Interpersonal Networks, and Attitude Polarization. *Journal of Communication, 67*(2), 256-281. doi: 10.1111/jcom.12288

Stroud, N. J. (2008). Media Use and Political Predispositions: Revisiting the Concept of Selective Exposure. *Political Behavior, 30*(3), 341-366. doi: 10.1007/s11109-007-9050-9

Sunstein, C. R. (2007). *Republic.com 2.0*. Princeton, N.J: Princeton University Press.

Takhteyev, Y., Gruzd, A., & Wellman, B. (2012). Geography of Twitter networks. *Social Networks, 34*(1), 73-81. doi: http://dx.doi.org/10.1016/j.socnet.2011.05.006

Vaccari, C., Valeriani, A., Barberá, P., Jost, J. T., Nagler, J., & Tucker, J. A. (2016). Of Echo Chambers and Contrarian Clubs: Exposure to Political Disagreement Among German and Italian Users of Twitter. *Social Media + Society, 2*(3), 2056305116664221. doi: doi:10.1177/2056305116664221

Wasserman, S., & Faust, K. (1994). *Social Network Analysis*. Cambridge: Cambridge University Press.

Wojcieszak, M. (2010). 'Don't talk to me': effects of ideologically homogeneous online groups and politically dissimilar offline ties on extremism. *New Media & Society, 12*(4), 637-655. doi: doi:10.1177/1461444809342775

Wojcieszak, M., & Mutz, D. C. (2009). Online Groups and Political Discourse: Do Online Discussion Spaces Facilitate Exposure to Political Disagreement? *Journal of Communication, 59*(1), 40-56. doi: 10.1111/j.1460-2466.2008.01403.x

Zollo, F., Bessi, A., Del Vicario, M., Scala, A., Caldarelli, G., Shekhtman, L., et al. (2017). Debunking in a world of tribes. *PLOS ONE, 12*(7), e0181821. doi: 10.1371/journal.pone.0181821

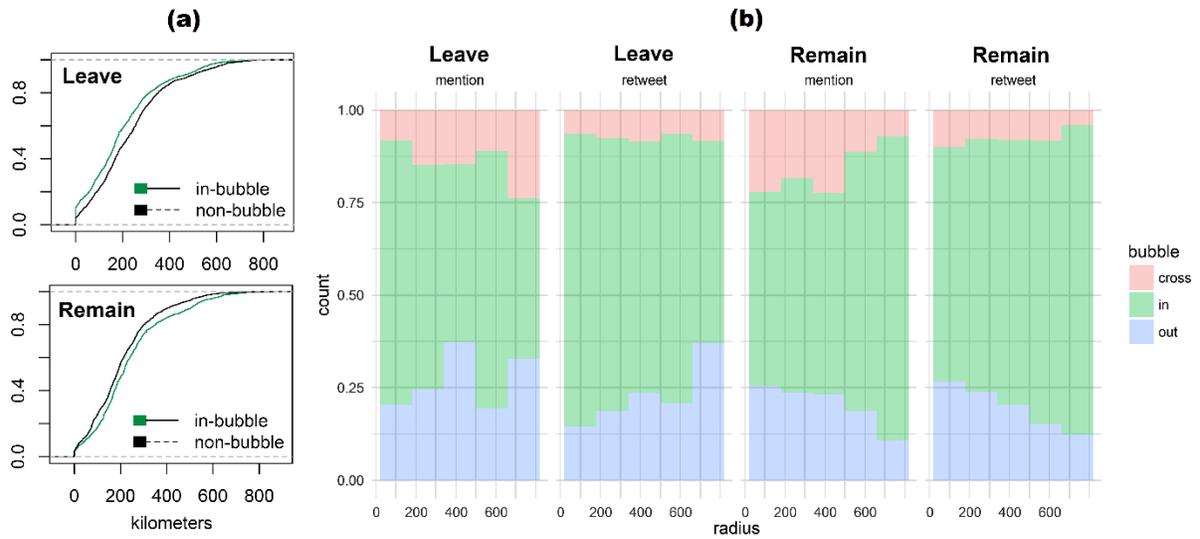

Figure 1: (a) Cumulative Distribution Function (CDF) of in-bubble (echo chambers) and non-bubble (out- and cross-bubble) communication; and (b) Histogram of distance travelled by messages between sender and receiver in 50-kilometer bins

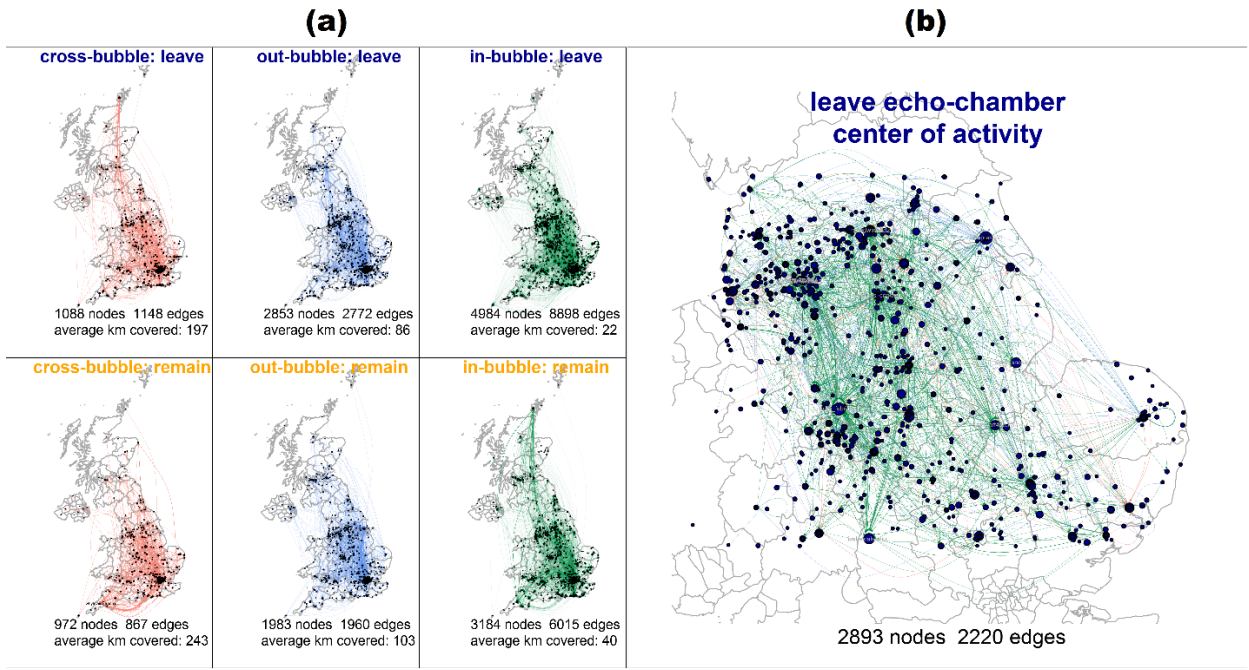

Figure 2: (a) Geographic pattern of cross-bubble, out-bubble, and echo chambers (in-bubble) with number of vertices and edges in each subgraph and the average distance travelled between sender and receiver; and (b) snapshot of the central point of diffusion of the Leave campaign, geographically located in the English Midlands, the North, and the East.

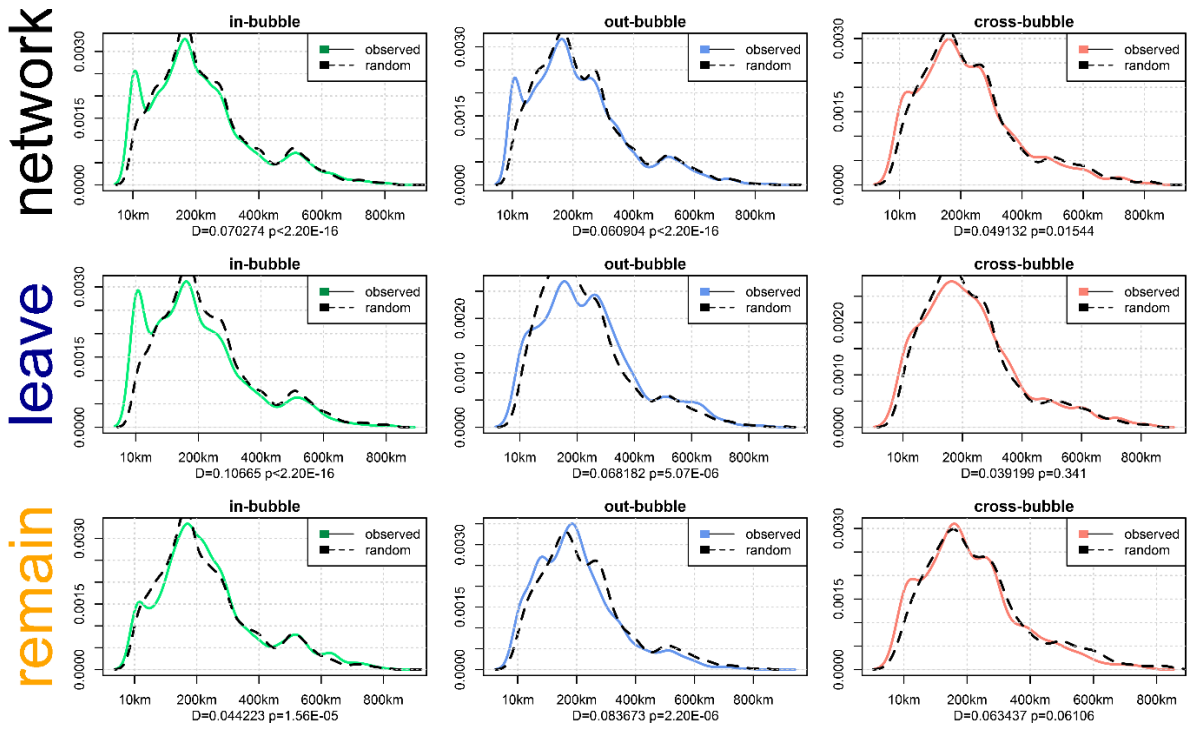

Figure 3: Distribution of distances covered by interactions across in-bubble, out-bubble, and cross-bubble for referendum network and subgraphs of the Leave and Remain campaigns, followed by the Kolmogorov–Smirnov test statistic and the *p*-value. The dotted line shows the reference probability distribution used to test the similitude of the two samples with a continuous distribution.

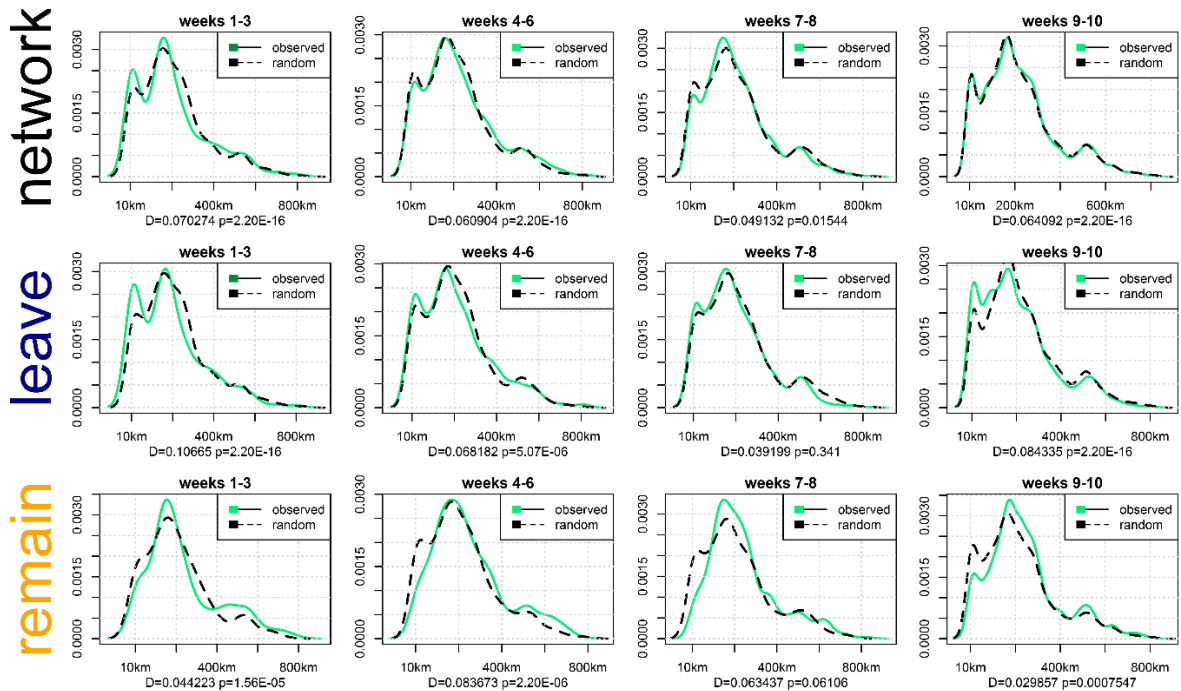

Figure 4: Distribution of distances covered by echo-chamber communication (in-bubble) for the referendum network and subgraphs of the Leave and Remain campaigns over the 10-week referendum campaign (14 April to 23 June). The Kolmogorov–Smirnov test statistic and the *p*-value indicate the differences in the observed and randomized distributions: while in weeks 9-10 observed and randomized signals are similar at the network level, there are remarkable and inversed interactions between online activity and geography in the Leave and Remain subgraphs.